\documentstyle[aps,epsf]{revtex}
\begin{document}
\draft
\preprint{gr-qc/0009008}
\title{Quantum Aspects of Black Hole Entropy\footnote{Plenary talk at the
International Conference on Gravitation and Cosmology 2000, Kharagpur,
India, January 2000.}}
\author{Parthasarathi Majumdar}
\address{The Institute of Mathematical Sciences, CIT Campus, Chennai
600113, India.\footnote{email: partha@imsc.ernet.in }}
\maketitle
\begin{abstract}
This survey intends to cover recent approaches to black hole entropy
which attempt to go beyond the standard semiclassical perspective.  
Quantum corrections to the semiclassical Bekenstein-Hawking area law for
black hole entropy, obtained within the quantum geometry framework, are
treated in some detail. Their ramification for the holographic entropy
bound for bounded stationary spacetimes is discussed. Four dimensional
supersymmetric extremal black holes in string-based $N=2$ supergravity are
also discussed, albeit more briefly.
\end{abstract}

\section{Introduction}
Despite the fact that gravitational collapse is a cataclysmic
phenomenon wherein a multitude of physical processes (some understood,
others yet to be discovered) are unleashed, the end-product -- a black hole
-- is a pristine object. As Chandrasekhar says, ``...the only
elements in the construction of black holes are our concepts of space and 
time. They are, thus, almost by definition, the most perfect macroscopic 
objects there are in the universe. And since the general theory of 
relativity provides a unique three-parameter family of solutions for 
their description, they are the simplest objects as well." \cite{chandra} 
But this does not tell the full story. The simplicity and perfection 
depicted in Chandrasekhar's description were dramatically challenged in
the early seventies by Jacob Bekenstein \cite{bek} and Steven Hawking
\cite{haw1}, based on considerations that germinate from the known quantum
origin of all matter (and radiation). 

Bekenstein started with the laws of black hole mechanics \cite{bch}, viz., 
\begin{itemize}
\item{\it The Zeroth Law} which states that the surface gravity of a black
hole is a constant on the event horizon.
\item{\it The First law} which, for the Kerr-Newman solution, states that
\begin{equation}
dM~\equiv~\Theta~d {\cal A}_h~+~\Phi~dQ~+~{\vec \Omega} \cdot
d{\vec L}~, ~\label{kern}
\end{equation}
where, $\kappa \equiv (r_+ - r_-)/4 {\cal A}_{hor}$ is the surface
gravity, $\Phi \equiv 4\pi Qr_+ /{\cal A}_{h}$ is the
electrostatic potential at the horizon and ${\vec \Omega}
\equiv 4\pi {\vec L}/M {\cal A}_{h}~$ is the angular velocity at the
horizon. 
\item{\it The Second law} which states that the area of the event horizon
of a black hole can never decrease. In other words, if two black holes of
horizon area ${\cal A}_1$ and ${\cal A}_2$ were to fuse, the area of the
resultant black hole ${\cal A}_{12} > {\cal A}_1+{\cal A}_2$. 
\end{itemize}

Observing the analogy of these laws with the laws of ordinary
thermodynamics, and gleaning insight from information theory, Bekenstein
made the bold proposal that a black hole must have an entropy
$S_{bh}$ proportional to the area of its horizon,
\begin{equation}
S_{bh}~~=~~const.~\times~{\cal A}_{h}~~.\label{entro}
\end{equation}
The constant in eq. (\ref{entro}) was fixed to be $1/4G$ where $G$ is
Newton's constant, by Hawking \cite{haw1} who also showed that a black
hole placed in a thermal background {\it cooler} than the `temperature'
given by the surface gravity (Hawking temperature) must radiate in a
thermal spectrum. This remarkable result also confirmed the Generalized
Second Law of thermodynamics \cite{bek}. 
 
The black hole entropy $S_{bh}$ arises from our lack of information about
the nature of gravitational collapse. The post-collapse configuration is
completely characterised by three parameters, viz., $M~,~Q~,~{\vec L}$
which encode in an unknown way the diverse set of events occurring during
collapse, just as a thermodynamic system is characterised by a few
quantities like pressure, volume, temperature etc. which encode the
microstates of the system. The microstates responsible for the entropy
should be quantum gravitational.  However, a complete quantum gravity
theory which serves the purpose is still not available, so the best one
can do is to consider extant proposals for such a theory. 

To summarize, the black hole entropy problem consists of identifying and
counting the underlying quantum states in an attempt to verify if the
semiclassical Bekenstein-Hawking Area Law (BHAL) does indeed hold. 
However, as eloquently argued by Carlip \cite{car1}, this need not
necessitate details of a proposal for quantum theory of gravitation. The
semiclassical origin of BHAL might mean that one would be able to deduce
it on the basis of some classical symmetry principle. 

Carlip's approach basically consists of identifying a two dimensional
Virasoro subalgebra of the algebra of generators of diffeomorphisms in 
a spacetime of any dimension with Lorentzian signature. There appear to be
several possible ways in which such a subalgebra may emerge from the
algebra of diffeomorphisms, given a set of boundary conditions valid at
the black hole horizon. The Virasoro algebra thus obtained is then
quantized by correspondence, and its primary states identified with the
microstates responsible for black hole entropy. Appealing to the Cardy
formula for the asymptotic degeneracy of these states, one obtains the
BHAL. 

The classical Poisson algebra of the spacetime diffeomorphism generators
is given, for spacetimes with boundaries, by 
\begin{equation}
\{ H(\xi_1) , H(\xi_2) \}_{PB}~=~H \left( \{\xi_1 , \xi_2 \} \right)~+~K(
\xi_1, \xi_2)~. \label{cpa}
\end{equation}
For solutions of the Einstein equations (i.e., configurations for which
the diffeomorphism constraints are satisfied in the bulk part of the
spacetime), the non-trivial part of the algebra receives contribution only
from the boundary of spacetime. It follows that the central term of the
algebra (\ref{cpa}) has contributions only from the boundary. Assuming the
existence of standard fall-off conditions at asymptopia (for
asymptotically flat spacetimes), one then concludes that the central term
of (\ref{cpa}) is non-trivial because of the specific boundary conditions
imposed at the black hole event horizon which plays the role of the inner
boundary. Carlip's boundary conditions are ostensibly for local
Killing horizons which do not appear to need a global timelike Killing
vector which goes to a constant at null infinity. For such boundary
conditions in any dimensional Lorentzian spacetimes, the algebra
(\ref{cpa}) can be shown to have a two dimensional Virasoro sub-algebra
corresponding to conformal diffeomorphisms on the `$r-t$' plane. The
central term of this algebra can be calculated in terms of the central
`charge" $c$. 

Corresponding to this classical Virasoro algebra is a quantum Virasoro
algebra; the microstates describing the horizon are identified with the
primary states of this quantum Virasoro algebra. The macroscopic black
hole is supposed to be described by primary states with arbitrarily high
conformal weights. The boundary conditions also enable one to relate the
central charge to the area (considered large in Planck units) of the
horizon. One now appeals to the Cardy formula for the asymptotic density
of these highly excited primary states
\begin{equation}
\rho ~\sim~\exp 2\pi~\left[ {c \over 6} \left( \Delta - {c \over 24}
\right) \right]^{1/2}~\label{card}
\end{equation}
with $\Delta = {\cal A}_h / 8\pi l_P^2~,~c = 3 {\cal A}_h /2\pi
l_P^2~,~l_P^2 =G$. It is
easy to see that
\begin{equation}
S_{bh}~\equiv~\ln \rho~=~S_{BH}~\label{carl}
\end{equation}
where, $S_{BH} \equiv {\cal A}_h/4l_P^2$. 

This implies that proposals for a quantum theory of gravity must have
predictions for black hole entropy {beyond} the BHAL. In other words,
there must be quantum corrections to the BHAL which cannot be deduced on
the basis of semiclassical reasoning alone. On dimensional grounds, one
expects that the {\it quantum corrected} black hole entropy might look
like
\begin{equation} 
S_{bh}~~=~~S_{BH}~+~\delta_q S_{bh} \label{cor}
\end{equation}
where,
\begin{equation}
\delta_q S_{bh}~=~\sum_{n=0}^{\infty} C_n~{\cal A}_h^{-n}~~\label{gen}
\end{equation}
where, ${\cal A}_h$ is the classical horizon area and $C_n$ are
coefficients which are independent of the horizon area but dependent on
the Planck length (Newton constant). However, in principle, one could
expect an additional term proportional to $ln~{\cal A}_h$ as the leading
quantum correction to the semiclassical $S_{BH}$. Such a term is expected
on general grounds pertaining to breakdown of na\"ive dimensional analysis
due to quantum fluctuations, as is common in quantum field theories in
flat spacetime and also in quantum theories of critical phenomena.

There is yet another restriction that one may want to subject these
quantum corrections to. This restriction originates from the so-called
Holographic principle and the ensuing entropy bound \cite{jw} -
\cite{smo}. According to this principle, the maximum entropy a spacetime
with boundary can have is bounded from above by the BHAL. If this
restriction is taken
seriously, it follows that $\delta_q S_{bh} \leq 0$. In the literature
there are calculations of the so-called Entanglement entropy. These are
based on quantum field theory in fixed classical black hole backgrounds
\cite{mann} and logarithmic corrections have been obtained. But these
corrections (a) appear to depend on some undetermined renormalization
scale, (b) appear to be in discord with the restriction due to the
holographic entropy bound and (c) they do not include fluctuations of the
background geometry. In the sequel, we consider two calculations of
post-BHAL corrections where spacetime fluctuations are taken into account.
One is the approach known as quantum geometry and the other based on $N=2$
supergravity arising out of type II string theory. While preferentially
dealing with the former approach in more detail, we shall also relate it
to a tightening of the entropy bound arising from the holographic
principle.
\section{Quantum Geometry}
This approach, also called Quantum General Relativity, envisages an exact 
solution to the problem of quantization of standard four dimensional general 
relativity, in  contrast to a perturbative expansion around a flat 
classical background. 
\subsection{Classical Connection Dynamics}
The canonical treatment of classical general relativity, otherwise known 
as geometrodynamics \cite{adm}, is traditionally formulated in terms 
of 3-metrics, i.e, restrictions of the metric tensor to three dimensional 
spacelike hypersurfaces (`time slices'). Canonically conjugate variables 
to these are then constructed, Poisson brackets between them defined and 
the entire set of first class constraints derived. The problem with this 
is that the constraints remain quite intractable. 

A significant departure from this approach is to formulate canonical
general relativity as a theory of `gauge' connections, rather than
3-metrics \cite{ash1}. Some of the constraints simplify markedly as a
consequence, allowing exact treatment, although this is not true for all
the constraints (e.g., the Hamiltonian constraint still remains difficult
to analyse). The method has also undergone substantial evolution since its
inception; a complex one-parameter family of connection variables is
available as one's choice of the basic `coordinate' degrees of freedom. 
The original Ashtekar choice \cite{ash1}, viz., the self-dual $SL(2,C)$
connection (inspired by work of Amitabha Sen \cite{titu}), corresponding
to one member of this family, is `geometrically and physically
well-motivated' because the full tangent space group then becomes the
gauge group of the canonical theory \cite{imm}. However, quantizing a
theory with complex configuration degrees of freedom necessitates the
imposition of subsidiary `reality' conditions on the Hilbert space,
rendering the formulation unwieldy. Also, given that the quantum wave
function is a functional of the connection, one is led to dealing with
complex functional analysis which is a difficult tool to use efficiently
to extract physical results. 

A better alternative, related to the former by canonical transformations, is
to deal with the Barbero-Immirzi family of {\it real} $SU(2)$ connections
confined to the time-slice M 
\begin{equation} 
A_i^{(\beta) a} ~~\equiv~~ \epsilon
^{abc} \Omega_i~_{bc} ~ +~ \beta g_{ij} \Omega^j~_{0a}~, \label{imcon}
\end{equation} 
labelled by a positive real number $\beta (\sim O(1))$ known
as the Barbero-Immirzi (BI) parameter \cite{imm}.\footnote{Here 
$\Omega_{\mu}~^{BC}$ is the standard Levi-Civita connection, $i,j = 1,2,3$ are spatial
world indices, and $a,b,c =1,2,3$ are spatial tangent space indices, and
$g_{ij}$ is the 3-metric.} This yields the BI family of curvatures
(restricted to M), 
\begin{equation} F^{(\beta) a}_{ij}
~~\equiv~~\partial_{[i} A_{j]}^{(\beta) a}~+~\epsilon^{abc} ~A_i^{(\beta) b}~
A_j^{(\beta) c}~. \label{curv} \end{equation} 
The variables canonically
conjugate to these are given by the so-called solder form 
\begin{equation}
E_i^{(\beta) a}~~=~~\frac{1}{\beta}~\epsilon^{abc} \epsilon_{ijk} e_j~^b
e_k~^c~.  \label{sold} \end{equation} 
The canonical Poisson bracket is then given by 
\begin{equation} 
\left \{ A_i^{(\beta) a} (x)~,~E_j^{(\beta) b}(y)
\right \}~=~\beta~ \delta_{ij}~\delta^{ab}~ \delta(x,y)~. \label{pb}
\end{equation}
\subsection{Quantization}
Canonical quantization in the connection representation implies that 
physical states are gauge invariant functionals of $A_i^{(\beta)a}(x)$ and 
\begin{equation}
E_i^{(\beta) a}~\rightarrow ~ {\hat E}_i^{(\beta) a}~\equiv~{{\beta \hbar} \over 
i}~{\delta \over {\delta A_a^{(\beta) i}}}~. \label{mom} \end{equation}
A useful basis of states for solution of the quantum constraints are the `spin 
network' states which generalize the loop space states used earlier 
\cite{rov1}. A spin network consists of a collection {\it edges} and {\it 
vertices}, such that, if two distinct edges meet, they do so in a vertex. It is 
a lattice whose edges need not be rectangular, and indeed may be non-trivially 
knotted. E.g., the graph shown in fig. 1 has 9 edges and 6 vertices. 
\begin{figure}%[htbp]
\epsfxsize=8cm
\centerline{\epsfbox{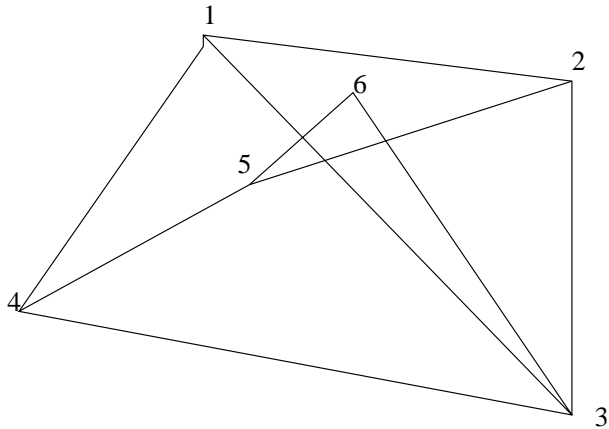}}
\caption{A spin network with 9 edges and 6 vertices}
\label{fig:fig1}
\end{figure}

To every edge $\gamma_l$ ($l=1,2, ..., 9$) we assign a spin $J_l$ which takes 
all half-integral values including zero. Thus, each edge transforms as a 
finite dimensional irreducible representation of $SU(2)$. In addition, one 
assigns to each edge a Wilson line functional of the gauge 
connection $h_l(A) = {\cal P}~\exp~\int_l~d\gamma^i_l~(A \cdot \tau)_i$, where 
$\tau^a$ are $SU(2)$ generators in the adjoint representation. To every vertex 
is assigned an $SU(2)$ invariant tensor $C^v$. These assignments completely 
define the basis states, which form a dense set in the Hilbert space of gauge 
invariant functionals of ${^{\beta}}A$. The inner product of these states then 
induces a measure on the space of connections which can be used to define a 
`loop transform' \cite{imm} of physical states, representing the same state, by 
diffeomorphism invariance. `Weave' states, supported on complicated and fine 
meshed nets  (with meshes of Planck scale size) are supposedly typical physical 
states. Thus, the classical spacetime continuum metamorphoses in the quantum 
domain into a space 
of `weaves' with meshes of Planck scale size on which all curvature (and indeed 
all dynamics) is concentrated. The Einsteinian continuum emerges when we view 
the weaves from afar, and are no longer able to see the meshes. 

Observables on the space of physical states (like the weaves) include 
geometrical operators like the area and volume operators, which typically are 
functionals of the canonical variables. To calculate the spectrum of these 
operators in the connection representation requires a technique of 
`regularization' since the classical definition of these quantities translates 
into singular objects upon naive quantization. E.g., the area operator
${\hat {\cal A}} (S)$ corresponding to a two dimensional surface $S$ 
intersecting a subset ${\cal L}$ of edges of a net, not touching any of its 
vertices and having no edge lying {\it on} $S$ is formally defined as 
\begin{equation}
{\hat {\cal A}}(S)~\psi_n~\equiv~ \left( ~\int d^2 \sigma \sqrt{n_i n_j {\hat 
E}^{ia} {\hat E}^{jb}}~ \right)_{reg}~ \psi_n~. \end{equation}
For large areas compared to $l_{Planck}^2$, this reduces to \cite{rov2}, 
\cite{ash3} 
\begin{equation}
{\hat {\cal A}}(S)~\psi_n~=~\beta \hbar l_{Planck}^2 ~\sum_{l \epsilon 
{\cal L}} \sqrt{J_l (J_l+1)}~\psi_n~~. \label{darea} \end{equation}
The discreteness in the eigen-spectrum of the area operator is of course 
reminiscent of 
discrete energy spectra associated with stationary states of familiar quantum 
systems. Each 
element  of the discrete set in (\ref{darea}) corresponds to a particular {\it 
number} of intersections (`punctures') of the spin net with the boundary
surface $S$. Diffeomorphism invariance ensures the irrelevance of the
locations of punctures. This will have important ramifications later.
\subsection{Schwarzschild black hole: boundary conditions}
We now consider the application of the foregoing formalism to the
calculation of entropy of the four dimensional Schwarzschild black hole,
following \cite{rov2}, \cite{car1}, \cite{kras}, \cite{ash4} and
\cite{km}. The basic idea is to concentrate on the horizon as a boundary
surface of spacetime (the rest of the boundary being described by the
asymptotic null infinities ${\cal I}^{\pm}$), on which are to be imposed
boundary conditions specific to the horizon geometry of the Schwarzschild
black hole vis-a-vis its symmetries etc. These boundary conditions then
imply a certain description for the quantum degrees of freedom on the
boundary. The entropy is calculated by counting the `number' of boundary
degrees of freedom. The region of spacetime useful for our purpose is the
`late time' part $\Delta$ of the event horizon ${\cal H}$ for which
nothing crosses ${\cal H}$, and it is of constant cross-sectional area
${\cal A}_S$. A particular spacelike foliation of spacetime is considered,
which intersects ${\cal H}$ (in particular $\Delta$) in the 2-sphere $S$.

Standard asymptotically flat boundary conditions are imposed on ${\cal 
I}^{\pm}$; those on the event horizon essentially subsume the following: 
first of all, the horizon is a null surface with respect to the 
Schwarzschild metric; second, the black hole is an isolated one with no 
gravitational radiation on the horizon; thirdly, the patch $\Delta$ has two flat 
(angular) coordinates spanning a special 2-sphere which {\it coincides} with 
$S$, the intersection of the time-slice $M$ with $\Delta$. The last 
requirement follows from the spherical symmetry of the Schwarzschild 
geometry. These boundary conditions have a crucial effect on the 
classical Hamiltonian 
structure of the theory, in that, in addition to the bulk contribution to 
the area tensor of phase space (the symplectic structure) arising in 
canonical general relativity, there is a {\it boundary} contribution. Notice 
that the boundary of the spacelike hypersurface M intersecting the black 
hole horizon is the 2-sphere $S$. Thus, the symplectic structure is given by
\begin{eqnarray}
\Omega|_{ A^{(\beta)}, E^{(\beta)} } (~ \delta E^{(\beta)}~,~\delta 
A^{(\beta)}~&;&~\delta E^{(\beta)'}~,~ \delta A^{(\beta)'} ~) 
\nonumber \\
&=&{1 \over {8\pi G}}~\int_{\em M}~Tr~( \delta E^{(\beta)} \wedge \delta 
A^{(\beta)'}~-~\delta E^{(\beta)'} \wedge \delta  A^{(\beta)} )~\nonumber \\
&-& {k \over {2\pi}}~\int_{S=\partial {\em M}}~ Tr ( ~\delta 
A^{(\beta)} \wedge \delta A^{(\beta)'} )~, \label{symp} \end{eqnarray}
where, $k \equiv {{\cal A}_S \over {2 \pi \beta G}}$. The second term in 
(\ref{symp}) corresponds to the boundary contribution to the symplectic 
structure; it is nothing but the symplectic structure of an $SU(2)$ level 
$k$ Chern Simons theory living on M. This is consistent with an extra term 
that arises due to the boundary conditions in the action, that is exactly an 
$SU(2)$ level $k$ Chern Simons action on $\Delta$ \cite{ash4}. As a
consequence of the boundary Chern Simons term, the curvature pulled back
to $S$ is proportional to the pullback (to $S$) of the solder form 
\begin{equation}
F^{(\beta)}~+~{2\pi \beta \over {\cal A}_S}~E^{(\beta)}~=~0~~.\label{cons} 
\end{equation}
This is a key relation for the entropy computation. 

The aforementioned boundary conditions are special cases of a larger class
which define the so-called Isolated horizon \cite{ash5} which includes
charged non-rotating back holes (including those in the extremal limit),
dilatonic black holes and also cosmological horizons like de Sitter space.
For the entire class, eq. (\ref{cons}) holds on the 2-sphere $S$
constituting the intersection of the spatial slice $M$ with $\Delta$. What
follows therefore certainly applies to non-rotating isolated horizons. 

\subsection{Quantum entropy calculation}

In the quantum theory, we have already seen that spacetime `in the bulk'
is described by spin nets ($\{ \psi_V \}$ say) at fixed time-slices.  It
has been shown \cite{ash3} that spin network states constitute an
eigen-basis for the solder form with a discrete spectrum. Now, in our
case, because of the existence of the event horizon which forms a boundary
of spacetime, there are additional surface states $\{ \psi_S \}$
associated with Chern Simons theory. In the canonical framework, the
surface of interest is the 2-sphere $S$ which forms the boundary of M. 
Thus, typically a state vector in the Hilbert space {\bf H} would consist
of tensor product states $\psi_V \otimes \psi_S$. Eq. (\ref{cons}) would
now act on such states as an operator equation. It follows that the
surface states $\{ \psi_S \}$ would constitute an eigenbasis for
$F^{(\beta)}$ restricted to $S$, with a discrete spectrum. In other words,
{\it the curvature has a support on $S$ only at a discrete set of points
-- punctures}. These punctures are exactly the points on $S$ which are
intersected by edges of spin network `bulk' states, in the manner
discussed earlier for the definition of the area operator. At each
puncture $p$ therefore one has a specific spin $J_p$ corresponding to the
edge which pierces $S$ at $p$. The black hole can then be depicted (in an
approximate sense) as shown in fig. 2. 

\begin{figure}%[htbp] 
\epsfxsize=8cm
\centerline{\epsfbox{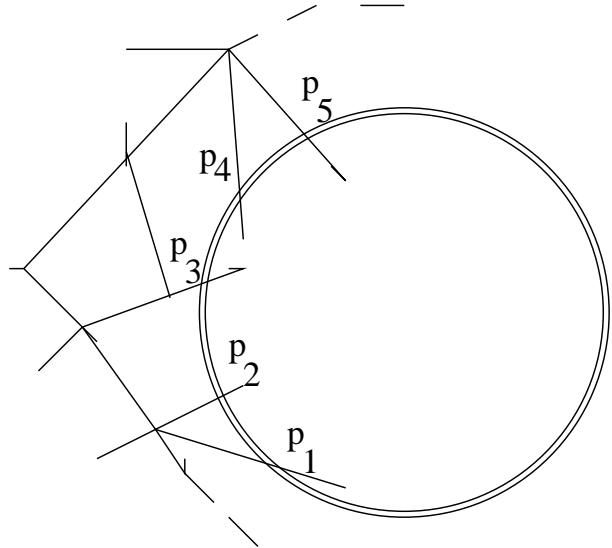}}
\caption{The bulk spin network intersecting the horizon at punctures}
\label{fig:fig2}
\end{figure}

Consider now a set of punctures ${\cal P}_{(n)} = \{ p_1,J_{p_1}~;~p_2,
J_{P_2}~;~ \dots p_n, J_{p_n} \}$.  For every such set, there is a
subspace ${\bf H}_V^{\cal P} $ of ${\bf H}_V$ which describes the space of
spin net states corresponding to the punctures. Similarly, there is a
subspace ${\bf H}S^{\cal P}$ of ${\bf H}_S$ describing the boundary Chern
Simons states corresponding to the punctures in ${\cal P}$. The full
Hilbert space is given by the direct sum, over all possible sets of
punctures, of the direct product of these two Hilbert (sub)spaces, {\it
modulo} internal gauge transformations and diffeomorphisms.\footnote{The
latter symmetry, in particular, as already mentioned, implies that the
location of punctures on $S$ cannot have any physical significance.} Now,
given that the Hamiltonian constraint cannot be solved exactly, one
assumes that there is at least one solution of the operator equation
acting on the full Hilbert space, for a given set of punctures ${\cal P}$. 

One now assumes that it is only the surface states $\psi_S$ that
constitute the microstates contributing to the entropy of the black hole
$S_{bh}$, so that the volume states $\psi_V$ are traced over, to yield the
black hole entropy as
\begin{equation}
S_{bh}~~=~~ln \sum_{\cal P}~dim~ {\bf H}_S^{\cal P}~~. \label{entr}
\end{equation}
The task has thus been reduced to computing the number of $SU(2)_k$ Chern
Simons boundary states for a surface with an area that is ${\cal A}_S$ to
within $O(l_{Planck}^2)$. One now recalls a well-known correspondence
between the dimensionality of the Hilbert space of the Chern Simons theory
and the number of conformal blocks of the two dimensional conformal field
theory (in this case $SU(2)_k$ Wess-Zumino-Witten model) `living' on the
boundary \cite{wit}. This correspondence now simplifies the problem
further to the computation of the number of conformal blocks of the WZW
model. Thus, {\it the problem of counting the microstates contributing to
the entropy of a 4d Schwarzschild black hole has metamorphosed into
counting the number of conformal blocks for a particular 2d conformal
field theory.}

This number can be computed in terms of the so-called fusion matrices
$N_{ij}^{~~r}$ \cite{dms}
\begin{equation}
N^{\cal P}~=~~\sum_{\{r_i\}}~N_{j_1 j_2}^{~~~~r_1}~ N_{r_1
j_3}^{~~~~r_2}~ N_{r_2
j_4}^{~~~~r_3}~\dots  \dots~ N_{r_{p-2} 
j_{p-1}}^{~~~~~~~~j_p} ~ \label{fun} \end{equation}
This is very similar to the composition of angular momentum in ordinary
quantum mechanics; it has been extended here to the infinite dimensional
affine Lie algebra $SU(2)_k$. Diagrammatically, this can be represented as
shown in fig. 3 below. 
\begin{figure}%[htbp]
\epsfxsize=8cm
\centerline{\epsfbox{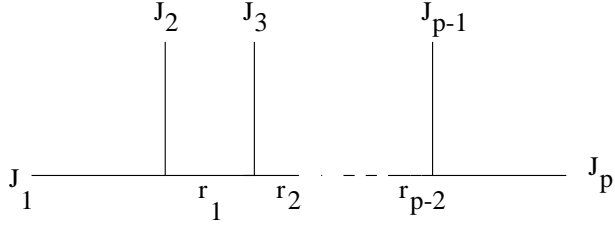}}
\caption{Diagrammatic representation of composition of spins $J_i$ for
$SU(2)_k$} 
\label{fig:fig3}
\end{figure}

Here, each matrix element $N_{ij}^{~~r}$ is $1 ~or~ 0$, depending on
whether the primary field $[\phi_r]$ is allowed or not in the conformal
field theory fusion algebra for the primary fields $[\phi_i]$ and
$[\phi_j] $ ~~($i,j,r~ =~ 0, 1/2, 1, ....k/2$):  
\begin{equation} 
[\phi_i]~ \otimes~ [\phi_j]~=~~\sum_r~N_{ij}^{~~r} [\phi_r]~ . \label{fusal}
\end{equation} 
Eq. (\ref {fun}) gives the number of conformal blocks with
spins $j_1, j_2, \dots, j_p$ on $p$ external lines and spins $r_1, r_2,
\dots, r_{p-2}$ on the internal lines.

We next take recourse to the Verlinde formula \cite{dms}
\begin{equation}
N_{ij}^{~~r}~= ~\sum_s~{{S_{is} S_{js} S_s^{\dagger r }} \over S_{0s}}~,
\label{verl}
\end{equation}
where, the unitary matrix $S_{ij}$ diagonalizes the fusion  matrix. Upon
using the unitarity of the $S$-matrix, the algebra (\ref{fun}) reduces to 
\begin{equation}
N^{\cal P}~=~ \sum_{r=0}^{k/2}~{{S_{j_1~r} S_{j_2~r} \dots S_{j_p~r}} \over 
(S_{0r})^{p-2}}~. \label{red} \end{equation}
Now, the matrix elements of $S_{ij}$ are known for the case under 
consideration ($SU(2)_k$ Wess-Zumino model); they are given by
\begin{equation}
S_{ij}~=~\sqrt{\frac2{k+2}}~sin \left({{(2i+1)(2j+1) \pi} \over k+2} \right )~, 
\label{smatr} \end{equation}
where, $i,~j$ are the spin labels, $i,~j ~=~ 0, 1/2, 1,  .... k/2$. Using this 
$S$-matrix, the number of conformal blocks for the set of 
punctures ${\cal P}$ is given by
\begin{equation}
N^{\cal P}~=~{2 \over {k+2}}~\sum_{r=0}^{ k/2}~{ {\prod_{l=1}^p sin \left( 
{{(2j_l+1)(2r+1) \pi}\over k+2} \right) } \over {\left[ sin \left( {(2r+1) \pi 
\over k+2} \right)\right]^{p-2} }} ~. \label{enpi} \end{equation}

In the notation of \cite{ash4}, eq. (\ref{enpi}) gives the dimensionality,
$dim ~{\cal H}^{\cal P}_S$, {\it for arbitrary area of the horizon $k$ and
arbitrary number of punctures}. The dimensionality of the space of states
${\cal H_S}$ of CS theory on three-manifold with $S^2$ boundary is then
given by summing $N^{\cal P}$ over all sets of punctures ${\cal P}, ~
N_{bh}~=~\sum_{\cal P} N^{\cal P}$. Then, the entropy of the black hole is
given by $S_{bh}~=~\log N_{bh}$. 

Observe now that eq. (\ref{enpi}) can be rewritten, with appropriate
redefinition of 
dummy variables and recognizing that the product can be written as a
multiple sum,
\begin{equation}
N^{\cal P}~=~\left ( 2 \over {k+2} \right) ~\sum_{l=1}^{k+1} sin^2 
\theta_l~\sum_{m_1 = -j_1}^{j_1} \cdots \sum_{m_p=-j_p}^{j_p} \exp \{
2i(\sum_{n=1}^p m_n)~ \theta_l \}~, 
\label{summ} \end{equation}
where, $\theta_l ~\equiv~ \pi l /(k+2)$. Expanding the $\sin^2 \theta_l$ and 
interchanging the order of the summations, a few manipulations then yield
\begin{equation}
N^{\cal P}~=~\sum_{m_1= -j_1}^{j_1} \cdots \sum_{m_p=-j_p}^{j_p} \left[ 
~\delta_{(\sum_{n=1}^p m_n), 0}~-~\frac12~ \delta_{(\sum_{n=1}^p m_n), 1}~-~ 
\frac12 ~\delta_{(\sum_{n=1}^p m_n), -1} ~\right ], \label{exct}
\end{equation}
where, we have used the standard resolution of the periodic Kronecker
deltas in terms of exponentials with period $k+2$,
\begin{equation}
\delta_{(\sum_{n=1}^p m_n), m}~=~ \left( 1 \over {k+2} \right)~
\sum_{l=0}^{k+1} \exp 
\{2i~[ (\sum_{n=1}^p m_n)~-~m] \theta_l \}~. \label{resol}
\end{equation}
Notice that the explicit dependence on $k+2$ is no longer present in the
exact formula (\ref{exct}).

The foregoing calculation does not assume any restrictions on $k$ or
number of punctures $p$. Eq. (\ref{exct}) is thus an exact formula
for the quantum entropy of a Schwarzschild black hole, or for that matter,
any non-rotating isolated horizon. 

Our interest focuses on the limit of large $k$ and $p$, appropriate to
macroscopic black holes of large area. Observe, first of all, that as $k
\rightarrow \infty$, the periodic Kronecker delta's in (\ref{resol}) 
reduce to ordinary Kronecker deltas \cite{dkm}, 
\begin{equation} 
\lim_{k \rightarrow \infty}~{\bar \delta}_{m_1+m_2+ \cdots +m_p,m}~=~
\delta_{m_1+m_2+ \cdots +m_p,m}~. \label{kinf} 
\end{equation} 
In this limit, the quantity $N^{\cal P}$ counts the number of $SU(2)$
singlet states, rather than $SU(2)_k$ singlets states. For a given set of
punctures with $SU(2)$ representations on them, this number is larger than
the corresponding number for the affine extension. 

Next, recall that the eigenvalues of the area operator for the horizon,
lying within one Planck area of the classical horizon area ${\cal A}_h$,
are
given by 
\begin{equation} 
{\hat {\cal A}}_h~\Psi_S~=~8\pi \beta 
~l_{P}^2~\sum_{l=1}^p~[j_l(j_l+1)]^{\frac12}~\Psi_S~, \label{area}
\end{equation} 
where, $l_{P}$ is the Planck length, $j_l$ is the spin on
the $l$th puncture on the 2-sphere and $\beta$ is the Barbero-Immirzi
parameter \cite{imm}. We consider a large fixed classical area of the
horizon, and ask what the largest value of number of punctures $p$ should
be, so as to be consistent with (\ref{area}); this is clearly obtained
when the spin at {\it each} puncture assumes its lowest nontrivial value
of 1/2, so that, the relevant number of punctures $p_0$ is given by
\begin{equation} 
p_0~=~{{\cal A}_h \over 4 l_{P}^2}~{\beta_0 \over \beta}~,\label{pmax} 
\end{equation} where, $\beta_0=1/\pi \sqrt{3}$. We are of course
interested in the case of very large $p_0$. 

Now, with the spins at all punctures set to 1/2, the number of states for
this set of punctures ${\cal P}_0$ is given by 
\begin{equation} 
N^{{\cal P}_0}~=~\sum_{m_1= -1/2}^{1/2} \cdots \sum_{m_{p_0}=-1/2}^{1/2}
\left[ ~\delta_{(\sum_{n=1}^{p_0} m_n), 0}~-~\frac12~ 
\delta_{(\sum_{n=1}^{p_0}
m_n), 1}~-~\frac12 ~\delta_{(\sum_{n=1}^{p_0} m_n), -1} ~\right ]
\label{excto} 
\end{equation} 
The summations can now be easily performed, with the result: 
\begin{equation} N^{{\cal P}_0}~=~\left( \begin{array}{c}
                         p_0 \\ p_0/2
                        \end{array} \right) 
                 ~ - ~\left( \begin{array}{c}
                         p_0 \\ (p_0/2-1) 
                         \end{array} \right)  ~\label{enpo} 
\end{equation}
There is a simple intuitive way to understand the result embodied in
(\ref{enpo}). This formula simply counts the number of ways of making
$SU(2)$ singlets from $p_0$ spin $1/2$ representations. The first term
corresponds to the number of states with net $J_3$ quantum number $m=0$
constructed by placing $m=\pm 1/2$ on the punctures.  However, this term
by itself {\it overcounts} the number of SU(2) singlet states, because
even non-singlet states (with net integral spin, for $p$ is an even
integer) have a net $m=0$ sector. Beside having a sector with total $m=0$,
states with net integer spin have, of course, a sector with overall $m=\pm
1$ as well. The second term basically eliminates these non-singlet states
with $m=0$, by counting the number of states with net $m=\pm 1$
constructed from $m=\pm 1/2$ on the $p_0$ punctures. The difference then
is the net number of $SU(2)$ singlet states that one is interested in for
that particular set of punctures. 

To get to the entropy from the counting of the number of conformal blocks,
we need to calculate $N_{bh}=\sum_{\cal P}~N^{\cal P}$, where, the sum is
over all sets of punctures. Then, $S_{bh}~=~ln N_{bh}$. 

The combination of terms in (\ref{enpo}), which corresponds to counting of
states in the $SU(2)$ Chern-Simons theory, yields a formula for the
quantum entropy of the black hole (isolated horizon). One can show that
\cite{ash3}, the contribution to $N_{bh}$ for this set of punctures ${\cal
P}_0$ with all spins set to 1/2, is by far the dominant contribution;
contributions from other sets of punctures are far smaller in comparison.
Thus, the entropy of an isolated horizon is given by the formula derived
in ref. \cite{km2}. Here, the formula follows readily from eq. 
(\ref{enpo}) and Stirling approximation for factorials of large integers. 
The number of punctures $p_0$ is rewritten in terms of area ${\cal A}_h$
through eq. (\ref{pmax}) with the identification $\beta~=~\beta_0~ln2$.
This allows us to write the entropy of an isolated horizon in terms of a
power series in horizon area ${\cal A}_h$: 
\begin{equation} 
S_{bh}~=~ {{\cal A}_h\over{4l_p^2}} ~-~{3\over
2}~ln \left( {{\cal A}_h \over {4l_p^2}} \right)~-~{1 \over 2}~ln
\left({\pi
\over{8(ln2)^3}} \right) ~-~O({\cal A}_h^{-1}).  \label{main} 
\end{equation}
Notice that the corrections to the BHAL are indeed finite and negative,
thereby conforming to the holographic constraint. 

Inclusion of representations other than spin $1/2$ on the punctures does
not affect the sign or coefficient of the logarithmic term.  For example,
we may place spin 1 on one or more punctures and spin $1/2$ on the rest.
The number of ways singlets can be made from this set of representations
can be computed in a straight forward way. Adding these new states to the
ones already counted above, merely changes the constant and order
${\cal A}_h^{-1}$ terms in formula (\ref{main}). The logarithmic
corrections are
therefore indeed robust. We may mention that very recently Carlip
\cite{car2} has presented compelling arguments that this formula may
possibly be of a universal character. 

\subsection{A tighter holographic entropy bound} 
The Holographic Principle \cite{jw} - \cite{suss} (HP) asserts that the
maximum possible number of degrees of freedom within a macroscopic bounded
region of space is given by a quarter of the area (in units of Planck
area) of the boundary. This takes into account that a black hole for which
this boundary is (a spatial slice of) its horizon, has an entropy which
obeys the Bekenstein-Hawking area law and also the generalized second law
of black hole thermodynamics \cite{bek1}. Given the relation between the
number of degrees of freedom and entropy, this translates into a
holographic Entropy Bound (EB) valid generally for spacetimes with
boundaries.\footnote{Spacetimes like the interior geometry of a black hole
or expanding cosmological spacetimes lie outside the purview of this
analysis, since even the Bekenstein bound is not valid for them
\cite{suss2}, \cite{bou}.}

The basic idea underlying both these concepts is a network, at whose
vertices are variables that take only two values (`binary', `Boolean' or
`pixel'), much like a lattice with spin one-half variables at its sites.
Assuming that the spin value at each site is {\it independent} of that at
any other site (i.e., the spins are {\it randomly} distributed on the
sites), the dimensionality of the space of states of such a network is
simply $2^p$ for a network with $p$ vertices. In the limit of arbitrarily
large $p$, such a network can be taken to approximate the macroscopic
surface alluded to above, a quarter of whose area bounds the entropy
contained in it. Thus, any theory of quantum gravity in which spacetime
might acquire a discrete character at length scales of the order of Planck
scale, is expected to conform to this counting and hence to the HP. 

Let us consider now a slightly altered situation: one in which the binary
variables at the vertices of the network considered are no longer
distributed randomly, but according to some other distribution. Typically,
for example, one could distribute them {\it binomially}, assuming, without
loss of generality, a large lattice with an even number of vertices. 
Consider now the number of cases for which the binary variable acquires
one of its two values, at exactly $p/2$ of the $p$ vertices. In case of a
lattice of spin 1/2 variables which can either point `up' or `down', this
corresponds to a situation of net spin zero, i.e., an equal number of
spin-ups and spin-downs. Using standard formulae of binomial
distributions, this number is
\begin{equation} N({\frac{p} {2}} |a) = 2^p~ \left( \begin{array}{c}
                         p \\ p/2
                        \end{array} \right)~[a~(1-a)]^{p/2} ~,
\label{bino} \end{equation}
Clearly, this number is maximum when the probability of occurrence
$a=1/2$;  it is given by $p! /(\frac{p}{2}!)^2$. Thus, the number of
degrees of freedom is now no longer $2^p$ but a smaller number. This
obviously leads to a lowering of the entropy. For very large $p$
corresponding to a macroscopic boundary surface, this number is
proportional to $2^p/p^{\frac12}$. The new EB can therefore be expressed
as 
\begin{equation} 
S_{max}~=~ln \left( {\exp{S_{BH}} \over S_{BH}^{1/2} } \right)~,
\label{newb} 
\end{equation} 
where, recall that $S_{BH}={\cal A}_h /4 l_P^2$ is the Bekenstein-Hawking
entropy. This is a tighter bound than that of ref.\cite{bek1} mentioned
above. We shall argue below that, in the quantum geometry framework, it is
possible to have an even tighter bound than that depicted in eq. 
(\ref{newb}).

Observe that eq. (\ref{enpo}) has two terms of which the second is
certainly negative. It follows that the number of $SU(2)$ singlet states
contributing to the entropy of the isolated horizon is bounded from above
by the first term
\begin{equation} N^{{\cal P}_0}~<~\left( \begin{array}{c}
                         p_0 \\ p_0/2
                        \end{array} \right)~.\label{u1b}
\end{equation} 
It may be pointed out that the rhs of (\ref{u1b}) also has another
interpretation. It represents the counting of boundary states for an
effective $U(1)$ Chern-Simons theory. It counts the number of ways unit
positive and negative $U(1)$ charges can be placed on the punctures to
yield a vanishing total charge. Upon using the Stirling approximation
for large $p_0$ and replacing it in terms of the classical horizon
area ${\cal A}_h$, this would then correspond to
an entropy bound given by the same formula (\ref{newb}) above for binomial
distribution of charges. 

But an even tighter bound on the entropy of the black hole (isolated
horizon) ensues from eq. (\ref{main}). It is obvious that this yields the
bound
\begin{equation} 
S_{max}~=~ln \left( {\exp{S_{BH}} \over S_{BH}^{3/2} }
\right)~. \label{newb1} 
\end{equation}
As already mentioned, this bound has a certain degree of robustness and
perhaps universality. 

The steps leading to the EB for any bounded spacetime now follows the
standard route of deriving the Bekenstein bound (see, e.g., \cite{smo}):
we assume, for simplicity that the spatial slice of the boundary of an
asymptotically flat spacetime has the topology of a 2-sphere on which is
induced a spherically symmetric 2-metric. Let this spacetime contain an
object whose entropy exceeds the bound. Certainly, such a spacetime cannot
have an isolated horizon as a boundary, since then, its entropy would have
been subject to the bound. But, in that case, its energy should be less
than that of a black hole which has the 2-sphere as its (isolated)
horizon. Let us now add energy to the system, so that it does transform
adiabatically into a black hole with the said horizon, but without
affecting the entropy of the exterior. But we have already seen above that
a black hole with such a horizon must respect the bound; it follows that
the starting assumption that the object, to begin with, had an entropy
violating the bound is not tenable.

One crucial assumption in the above arguments is that matter or
radiation crosses the isolated horizon
adiabatically in small enough amounts, such that the {\it isolated}
character of the horizon is not seriously affected. This is perhaps
not too drastic an
assumption. Thus, for a large class of spacetimes, one may propose
Eq.(\ref{newb1}) as the new holographic entropy bound. 

\section{Corrections to BHAL in 4d $N=2$ supergravity}
\subsection{Supersymmetric black holes}
Black hole solutions of four dimensional $N=2$ supergravity have been
studied extensively \cite{fer}, following improved understanding of the
Ramond-Ramond sector of string theories as the {\it solitonic} sector.
Likewise, these solutions lend themselves to a solitonic interpretation,
interpolating between two $N=2$ supersymmetric ground states which are
asymptotically flat geometries which behave like Bertotti-Robinson (e.g.,
$AdS_2 \otimes S_2$) geometries at the horizon. These solutions exhibit
only an $N=1$ supersymmetry globally, so that they saturate the
Bogomol'nyi-Prasad-Sommerfield (BPS) bound which relates the mass of the
black hole to its charge(s):
\begin{equation}
M_{ADM/Bondi}~\geq~G^{-1/2}~\left( Q^2 ~+~ {\tilde Q}^2 \right)^{1/2}~.
\label{bps}
\end{equation}

Clearly, these solutions represent {\it extremal} black holes. Almost a
decade earlier, in a pioneering paper Gibbons and Hull had shown that the
black hole configurations which saturated the BPS bound also led to the
existence of a chiral spin-half field (Killing spinor) which satisfied the
equation 
\begin{equation} 
D_{\mu} \epsilon(x)~=~\frac14~F_{\lambda \sigma} \gamma^{\lambda}
\gamma^{\sigma} \gamma_{\mu} \epsilon(x)~.\label{kils}
\end{equation} 
Observe that the spinor in question is the zero mode corresponding to
$N=1$ supersymmetry transformations and, as such, {\it defines} the notion
of a supersymmetric black hole. Gibbons and Hull also showed that, for
time independent Killing spinors, the {\it only} solutions saturating the
BPS bound are the Majumdar-Papapetrou class of solutions \cite{sdm}, given
by
\begin{eqnarray}
ds^2~&=&~-W^2~dt^2~+~W^{-2}~d{\bf x}^2~ \nonumber \\
F~&=&~\cos {\theta}~ F^0~+~\sin {\theta}~ {\tilde F}^0~\label{mp}
\end{eqnarray}
where,
\begin{eqnarray}
F^0~&=&~dW \wedge dt~\nonumber \\
W^{-1}~&=&~1~+~\sum_{s=1}^n~{M_s \over {|{\bf x} - {\bf
x}_s|}}~\label{defi}
\end{eqnarray}
The solution (\ref{mp}) represents an assembly of $n$ extremal
Reissner-Nordstrom black holes, each having a mass $M_s$ saturating the
bound (\ref{bps})  and identical electric/magnetic charges. In this
system, there is an exact cancellation of the gravitational attraction and
electrostatic repulsion, thereby guaranteeing stability. What Gibbons and
Hull's work clearly demonstrates is that the underlying reason for this
stability is the unbroken $N=1$ supersymmtry of the solution, defined as
above in terms of Killing spinors.

It is easy to check that the surface gravity of the assembly of extremal
black holes vanishes at the horizon of each individual black hole. The
assembly is thus stable against superradiance and Hawking radiation.
However, the solitonic character of the solution (\ref{mp}) becomes clear
only when it is coupled to $n$ $N=2$ vector multiplets carrying
Ramond-Ramond (electric/magnetic) charge. This coupled system
interpolates between asymptotpia and Bertotti-Robinson geometries which
are $N=2$ supersymmetric vacua.

\subsection{Entropy calculation}
The four dimensional $N=2$ supergavity multiplet consists of a Weyl
multiplet (which is itself composed of the graviton, two gravitinos), an
auxiliary multiplet containing complex scalar fields ${\tilde A}$ and also
superconformal gauge fields. Each $N=2$ vector multiplet contains a gauge
field, gauginos and a complex scalar field $X_s$. Together with ${\tilde
A}$, the $X_s$ constitute the moduli fields whose vacuum expectation
values determine the moduli space of BPS-saturating supersymmetric black
hole solutions. The coupling between the Weyl and vector multiplets is
described by the holomorphic function $F(X_s, {\tilde A})$ which is
protected from perturbative quantum corrections. As a result, it has been
shown that the effective action of $N=2$ supergravity is not renormalized
for such supersymmetric black hole backgrounds. 

In addition, the geometry of the moduli space exhibits a sort of
`fixed-point' behaviour: the moduli $X_s$ evolve into constants at the
horizon {\it regardless} of their value at asymptopia. Furthermore, these
constants are determined as a function of the Ramond-Ramond charges
parametrizing the black hole solution. The macroscopic (semiclassical) 
entropy of these configurations is then determined from the horizon area
which is expressed in terms of the constant values of the moduli by
inverting the formula relating them to the charges. For a single extremal
Reissner-Nordstrom black hole, this is simply 
\begin{equation}
S_{BH}~=~{\pi \over 16G}~|C|^2~\label{rn}
\end{equation}
where, $C$ is the fixed point value of the modulus field $X$. 

Now, $N=2$ supergravity is the low energy limit of type II superstring
theory, compactified to ${\cal M}_4 \otimes (CY)_3$ where, ${\cal M}_4$ is
Minkowski 4-space and $(CY)_3$ is three dimensional Calabi-Yau manifold.
The black hole solution extremizes the Born-Infeld effective action
corresponding to the Ramond-Ramond sector of the compactified string. This
sector has a quantized description in terms of open string states
terminating on D-branes propagating in a flat non-compact
transverse background spacetime. The black hole is taken to correspond (in
the weak coupling limit of the theory) to a set of BPS-saturating excited
string states with appropriate charge quantum numbers. The logarithm of
the degeneracy of these states, computed via the Cardy formula \cite{str} 
is defined as the {\it microscopic} entropy of the black
hole configuration, and agrees, for large values of the charges, to the
BHAL as depicted in (\ref{rn}). This remarkable agreement is attributed to
the non-renormalization of the effective action - a property of
supersymmetric theories. Thus, the microscopic entropy remains unchanged
as the coupling increases from weak to strong, the latter being identified
with the black hole domain. 

Now, one recalls that the low energy limit of string theory contains
general relativity (as a coupled spin two theory in a flat background),
together with corrections that can be represented as higher order
curvature terms in the classical action. These `stringy' corrections are
supposed to lead to corrections in the BHAL \cite{dew}. The basic tool
used here is the formalism due to Wald et. al. \cite{wald} in which black
hole entropy (in any diffeo-invariant theory) is represented as a Noether
charge corresponding to diffeomorphism invariance. There are two basic
assumptions in the approach followed in \cite{dew}: (a) there exist
BPS-saturating supersymmetric extremal black hole solutions of the
string-modifed higher curvature theory and (b) the moduli space of such
solutions exhibits the same fixed point behaviour as for the theory
without the higher curvature terms. There is a further important caveat: 
Wald's formalism requires that the Killing horizon be bifurcate, a
property that requires the surface gravity at the horizon to be
non-vanishing. On the other hand, extremal black holes have vanishing
surface gravity and degenerate horizons. This subtlety needs to be
addressed in a more careful treatment of the problem. The result obtained
so far can be expressed as 
\begin{equation} S_{bh}~=~S_{BH} ~-~{\pi \over 8}~Im \left( \partial F
\over \partial {\tilde A} \right)~.\label{correc}
\end{equation} The correction term on the rhs is to be evaluated at the
`fixed point', i.e., on the horizon. It turns out again that the result is
in `perfect' agreement with a `microscopic' calculation based on counting
of states in string theory in a flat background in weak coupling.

Several remarks regarding (\ref{correc}) are in order: (i) the
correction has
no obvious geometrical interpretation in terms of the horizon area; (ii) 
it is not at all clear whether the correction is positive or negative, so
that its status vis-a-vis the holographic restriction is uncertain.
Therefore, it is difficult to glean from this any implication for the
holographic entropy bound. A more disturbing aspect is the following: the
approach depends crucially on the `special' geometry of moduli space
which, in turn, is wholly dependent on the unphysical requirement of
unbroken supersymmetry. This means that generic non-extremal black holes,
of the type that most likely exist in the universe, are outside the
purview of any of these considerations.  

\section{Conclusions} 
The common technical tool, in both the quantum geometry and the string
theoretic approaches to counting black hole microstates, is two
dimensional conformal field theory. Whether this implies an unknown  
deeper connection, or merely exhibits a technical device {\it par
excellence} is yet to be ascertained. The quantum geometry approach
provides the most direct route to the relevant 2d CFT which, in the
foregoing, is a rational conformal field theory, viz., $SU(2)$ WZW model.
The string route has more ambiguities, although there are claims that
there could be a way to obtain the same answer for the quantum entropy
\cite{car2}. This latter aspect needs to be better understood. It is not
at all clear how two approaches so disparate in their premise can, even in
principle, lead to identical results.


\begin{thebibliography}{99}
\bibitem{chandra} S. Chandrasekhar, {\it Truth and Beauty}, pp 153-154, 
Chicago (1987). 
\bibitem{bek} J. Bekenstein, Phys. Rev. {\bf D7}, 2333 (1973).
\bibitem{haw1} S. Hawking, Comm. Math. Phys. {\bf 43}, 190 (1975).
\bibitem{bch} J. Bardeen, B. Carter and S. Hawking, Comm. Math. Phys.,
{\bf 31}, 161 (1973).
\bibitem{car1} S. Carlip, Class. Quant. Grav. {\bf 16}, 3327 (1999). 
\bibitem{jw} J. Wheeler, {\it It from bit}, Sakharov Memorial Lecture on
Physics, vol. 2, L. Keldysh and V. Feinberg, Nova (1992). 
\bibitem{thf} G. 't Hooft, {\it Dimensional reduction in quantum gravity}
gr-qc/9310006 in {\it Salam festschrift}, eds A. Alo, J. Ellis and S.
Randjbar-Daemi, World Scientific (1993). 
\bibitem{suss} L. Susskind, J. Math. Phys. {\bf 36}, 6377 (1995).
\bibitem{bek1} J. Bekenstein, Phys. Rev. {\bf D9}, 3292 (1974).  
\bibitem{smo} For a comprehensive review, see L. Smolin, {\it The strong
and weak holographic principles}, hep-th/0003056 and references therein. 
\bibitem{mann} R. Mann and S. Solodukhin, {\it Nucl. Phys.} {\bf 523B},
293 (1998) and references therein. 
\bibitem{adm} R. Arnowitt, S. Deser and C. Misner in {\it Gravitation: An 
Introduction to Current Research}, ed. L. Witten, Wiley (1967). 
\bibitem{ash1} A. Ashtekar, {\it Lectures on Non-perturbative Quantum  
Gravity}, (World Scientific, 1991).
\bibitem{titu} Amitabha Sen, Phys. Lett. {\bf 119B}, 89 (1982).
\bibitem{imm} F. Barbero, Phys. Rev. {\bf D51} (1995) 5507; G. Immirzi,  
{\it Nucl. Phys. Proc. Suppl.} {\bf 57}, 65 (1997). 
\bibitem{rov1} C. Rovelli, {\it Loop Quantum Gravity}, gr-qc/9710008.
\bibitem{rov2} C. Rovelli and L. Smolin, Phys. Rev. {\bf D52}, 5743
(1995); S. Fritelli, L. Lehner, C. Rovelli, Class. Quant. Grav. {\bf
13}, 2921 (1996).
\bibitem{ash3} A. Ashtekar and J. Lewandowski, Class. Quant. Grav. {\bf
14}, 55 (1997) and references therein.
\bibitem{kras} K. Krasnov, {\it On Quantum Statistical Mechanics of a
Schwarzschild black hole}, gr-qc/9605047, {\it Quantum geometry and
thermal radiation from black holes}, gr-qc/9710006.
\bibitem{ash4} A. Ashtekar, J. Baez, A. Corichi and K. Krasnov, Phys.
Rev. Lett. {\bf 80}, 904 (1998); A. Ashtekar, J. Baez and K. Krasnov, {\it
Quantum Geometry of Isolated  Horizons and Black Hole Entropy}, 
gr-qc/0005126.
\bibitem{km} R. Kaul and P. Majumdar, Phys. Lett. {\bf B439}, 267
(1998);  P. Majumdar, Ind. Jou. Phys. {\bf B43}, 147 (1998); R. Kaul,
{\it Topological field theories - a meeting ground for physicists
and mathematicians,} in {\it Quantum Field Theory: A 20th Century
Profile}, pp 211; ed. A.N. Mitra, published by Hindustan Book Agency
(India) and Indian National Science Academy, New Delhi (2000). 
\bibitem{ash5} A. Ashtekar, C. Beetle and S. Fairhurst, {\it Mechanics of
isolated horizons}, gr-qc/9907068, and references therein. 
\bibitem{wit} E. Witten, Comm. Math, Phys. {\bf 121}, 351 (1989).
\bibitem{dms} P. Di Francesco, P. Mathieu and D. Senechal, {\it Conformal
Field Theory}, p. 375 {\it et seq}  (Springer Verlag 1997).
\bibitem{dkm} S. Das, R. Kaul and P. Majumdar, {\it A new holographic
entropy bound from quantum geometry}, hep-th/0006211, to appear in Phys.
Rev. D.
\bibitem{km2} R. Kaul and P. Majumdar, Phys. Rev. Lett. {\bf 84}, 5255
(2000).
\bibitem{car2} S. Carlip, {\it Logarithmic Corrections to Black Hole
Entropy from Cardy Formula}, gr-qc/0005017. 
\bibitem{fer} S. Ferrara, R. Kallosh and A. Strominger, Phys. Rev. {\bf
D52}, 5412 (1995); A. Strominger, Phys. Lett. {\bf B383}, 39 (1996).
\bibitem{sdm} S. D. Majumdar, Phys. Rev. {\bf 72}, 390 (1947); A.
Papapetrou, Proc. Roy. Ir. Ac. {\bf A51}, 191 (1947).
\bibitem{str} J. Maldacena, A. Strominger and E. Witten, Jou. High. En.
Phy. {\bf 12}, 2 (1997).
\bibitem{suss2} W. Fischler and L. Susskind, {\it Holography and
Cosmology}, hep-th/9806039.  
\bibitem{bou} R. Bousso, Class. Quant. Grav. {\bf 17}, 997 (2000) and
references therein. 
\bibitem{dew} B. de Wit, {\it Modifications of the area law and $N=2$
supersymmetric black holes}, hep-th/9906095 and references therein.
\bibitem{wald} R. Wald, Phys. Rev. {\bf D48}, 3427 (1993); V. Iyer and R.
Wald, Phys. Rev. {\bf D50}, 846 (1994).

\end{thebibliography}
\end{document}